# Algorithmic Constitutionalism*

## Oren Perez & Nurit Wimer

ABSTRACT

The increasing encroachment of artificial intelligence (AI) on social life raises various risks to society, most prominently in the info-spheres created and controlled by Google, Facebook, Apple, and Amazon. We examine these risks through an in-depth discussion of the Facebook content moderation regime, which is already partially controlled by algorithms. We argue that the idea of ethical engineering, which was developed by the literature as a solution to the challenge of governing AI, is inadequate for various reasons. In this article, we develop a different approach to coping with the risks of AI governance, which we have termed "algorithmic constitutionalism." Our approach rests on three pillars: (a) layered architecture that consists of two levels of code: (i) operative or object level and (ii) meta level, whose purpose is to shield the core principles of the system from algorithmic-initiated changes; (b) algorithmic meta-reasoning, which allows the system to operate simultaneously at the two levels, so that it can (self) monitor, verify, and potentially correct in real time operations at the object level, if they depart from the principles protected by the meta-code level; and (c) correction by deliberation. We elaborate the idea of algorithmic constitutionalism and demonstrate how it can be applied to the Facebook content moderation regime. As part of this elaboration, we also consider the tension between societal and algorithmic constitutionalism. Paradoxically, the attempt to subject the AI algorithm to external deliberative control also opens the door for the AI agent to intervene in that process, potentially undermining its very purpose. We conclude by exploring the implications of our argument for the new European Digital Services Act, which came into force in October 2022.

* We would like to thank Angelo Golia and Gunther Teubner for their comments on an earlier draft of this paper.

The increasing encroachment of artificial intelligence (**AI**) on social life raises various risks to human welfare and human rights. There is a growing literature arguing that a prepotent AI **(PAI**)[1] may present an existential risk to humanity.[2] We argue that a focus on PAI shifts our attention away from the risks that existing hybrid human-digital systems present to human society today. These risks are most prominently found in the info-spheres created and controlled by Google, Facebook, Apple, and Amazon. Although these risks may not be existential, they are sufficiently serious to warrant close scrutiny.[3] We examine these risks through an in-depth discussion of Facebook governance system, which is already partially controlled by algorithms that perform various tasks, such as content moderation and classification, friend recommendation, ad delivery, and more.[4]

Second, we argue that ethical engineering, one of the primary approaches to the risk of PAI, does not offer a satisfactory solution to the challenge of governing AI.[5] The first difficulty with ethical engineering concerns moral pluralism.[6] Despite humanity's long history of moral inquiry, we lack a firm understanding of what the "right" moral theory is. The problem of moral pluralism becomes more acute as the range of problems AI is expected to resolve becomes broader.[7] A second shortcoming of the ethical engineering approach is the fact that the interaction of an AI agent with the external environment and the knowledge it may acquire through that process, coupled with a capacity for self-renewal, may lead it to develop new objectives and new interpretations of existing ones, which may be inconsistent with the ethical principles that the human designer has intended to implant in it.[8]

[1] We borrow the term from Critch Andrew & Krueger David., *Ai Research Considerations for Human Existential Safety (Arches),* (2020); other authors use the term "super-intelligent AI" or simply "super AI," see, e.g., Kovač, Mitja., *Judgement-Proof Robots and Artificial Intelligence: A Comparative Law and Economics Approach* (Mitja Kovač eds., 2020).
[2] Critch Andrew & Krueger David, supra note 1; BOSTROM NICK, SUPERINTELLIGENCE: PATHS, DANGERS, STRATEGIES (Oxford University Press, 2014); Yampolskiy Roman V., *Taxonomy of pathways to dangerous artificial intelligence*, *in* WORKSHOPS AT THE THIRTIETH AAAI CONFERENCE ON ARTIFICIAL INTELLIGENCE (2016). Paper presented at the Workshops at the thirtieth AAAI conference on artificial intelligence.
[3] Miller Michael L. & Vaccari Cristian, *Digital Threats to Democracy: Comparative Lessons and Possible Remedies*, *25*(3) THE INTERNATIONAL JOURNAL OF PRESS/POLITICS, 333 (2020); VAIDHYANATHAN SIVA, ANTISOCIAL MEDIA: HOW FACEBOOK DISCONNECTS US AND UNDERMINES DEMOCRACY (Oxford University Press, 2018).
[4] Ali Muhammad, Sapiezynski Piotr, Korolova Aleksandra, Mislove Alan, & Rieke Aaron, *Ad delivery algorithms: The hidden arbiters of political messaging* (2019) Available at: https://arxiv.org/pdf/1912.04255; Sean MacAvaney, Hao-Ren Yao, Eugene Yang, Katina Russell, Nazli Goharian & Ophir Frieder, *Hate Speech Detection: Challenges and Solutions,* 14(8) PLOS ONE (2019).
[5] Damien Trentesaux and Stamatis Karnouskos, *Engineering ethical behaviors in autonomous industrial cyber-physical human systems*, 24 COGNITION TECHNOLOGY & WORK, 113 (2022.
[6] Gordon John-Stewart, *Building Moral Robots: Ethical Pitfalls and Challenges*, 26.1 SCIENCE AND ENGINEERING ETHICS, 141 (2020).
[7] For example, the moral tradeoffs that underlie the moderation of speech on social media platforms is much more complex than those facing the control of autonomous vehicles. J-F. Bonnefon, A. Shariff & I. Rahwan, *The Trolley, The Bull Bar, and Why Engineers Should Care About the Ethics of Autonomous Cars [point of view]*, vol. 107, no. 3 PROCEEDINGS OF THE IEEE, 502 (2019); Luetge C., *The German Ethics Code for Automated and Connected Driving*, 30 PHILOS. TECHNOL, 547 (2017).

In this article, we develop a different approach to coping with the risks presented by algorithmic governance, which we have termed "algorithmic constitutionalism" **(AC**). Our approach rests on three pillars: (a) **Layered architecture** that consists of two levels of code: (i) operative or object level and (ii) meta level. The purpose of this layered architecture is to shield the core principles of the system (which are located at the meta-code level) from algorithmically initiated changes that can radically transform the system; (b) **Algorithmic meta-reasoning**, which allows the system to operate simultaneously at the two levels, so that it can (self) monitor, verify, and potentially correct in real time, operations at the object level, if they depart from the principles protected by the meta-code level. The meta-code can be figuratively described, following Giacomo et al., as a "restraining bolt;"[9] and (c) **Correction by deliberation**. Considering both the problem of moral pluralism and the risks that may emanate from an autonomous AI algorithm that has the power to rewrite the core principles of the system, we propose to limit the ability of the meta-level code to initiate corrective actions, by subjecting it to hardwired deliberation procedures. As we elaborate below, this proposal implies a departure from the optimization principle that underlies the dominant approach in AI development. Our thesis builds on humanity's long experience in taming power. Rather than relying on ethics and morality to guide it, human society has put its trust in politics, drawing on various decision-making procedures (from voting to polls), embedded in an intricate institutional structure of checks and balances. Together, these three principles form the basis of what we call "algorithmic constitutional law."

In the sections below we develop the idea of algorithmic constitutionalism and explore the legal, political, and technological challenges it faces while considering the connection between algorithmic and societal constitutionalism (SC). Our argument shares with SC the claim that state constitutions do not exhaust the universe of constitutionality, but it departs from SC by developing an algorithmic interpretation of constitutionality, which challenges the societal aspect of SC. We believe that one of the greatest challenges facing modern constitutional theory concerns the relation between algorithmic and societal constitutionalism. In the last two sections of the article, we further explore the tension between these two ideas.

---

[8] Costantini Stefania, *Ensuring trustworthy and ethical behaviour in intelligent logical agents*, 32.2 JOURNAL OF LOGIC AND COMPUTATION 443 (2022).

[9] See De Giacomo, Giuseppe, Luca Iocchi, Marco Favorito & Fabio Patrizi, *Foundations for Restraining Bolts: Reinforcement Learning With LTLf/LDLf Restraining Specifications*, vol. 29 PROCEEDINGS OF THE INTERNATIONAL CONFERENCE ON AUTOMATED PLANNING AND SCHEDULING, 128 (2019) and Costantini Stefania, *id*.

To make the discussion more concrete, we examine a hypothetical scenario in which Facebook becomes fully controlled by a semi-PAI. The Facebook autonomous platform law, with its myriad company statements, policies, internal standards, and algorithms constitutes a perfect arena for analyzing the tension between algorithmic and societal constitutionalism.[10] We consider in this context the recently approved EU Digital Services Act (**DSA**), which establishes a new regulatory regime for social media platforms, significantly reducing their capacity for self-regulation.[11]

The paper proceeds as follows. Section A discusses the risks associated with PAI. Section B examines the intersection between human agency and AI in the governance of Facebook (focusing on the domain of content moderation). Section C discusses the risks of removing humans from the Facebook compliance system and making it entirely AI-controlled. Section D describes our concept of algorithmic constitutionalism and applies it to the realm of Facebook. Section E concludes with a discussion of algorithmic and societal constitutionalism, drawing also on the new regulatory regime of the DSA.

## A. The risk of prepotent AI and the limits of ethical engineering

The idea that a PAI system may present an existential risk to humanity has received significant attention in the academic literature in the past few years. Nick Bostrom has defined the idea of PAI as "an intellect that is much smarter than the best human brains in practically every field, including scientific creativity, general wisdom and social

---

[10] Gorwa Robert, *What is Platform Governance? 22*(6) INFORMATION, COMMUNICATION & SOCIETY, 854 (2019); Owen Taylor, *Introduction: Why Platform Governance*? CENTRE FOR INTERNATIONAL GOVERNANCE INNOVATION (Oct 28, 2019). Available at: https://www.cigionline.org/articles/introduction-why-platform-governance; Timothy Garton Ash et al., supra note 4; Douek Evelyn, *Governing Online Speech*, 121(3) COLUMBIA LAW REVIEW, 759 (2021); Kadri Thomas E. & Klonick Kate, *Facebook v. Sullivan: Public Figures and Newsworthiness in Online Speech*, 93 S. CAL. L. REV., 37 (2019).

[11] Regulation (EU) 2022/2065 of the European Parliament and of the Council of 19 October 2022 on a Single Market For Digital Services and amending Directive 2000/31/EC (Digital Services Act).

skills."[12] He argued that if PAI evolved, humanity would most probably not be able to stop it.[13]

Bostrom emphasized that PAI has several unique properties that make it distinctive compared to other novel technologies, including (a) autonomy manifest in the ability to make independent decisions and take initiatives; (b) the development of distinctive motivational tendencies (not shared by humans); (c) capacity to self-replicate and self-improve; and (d) unique cognitive architecture which may guard AI agent against some kinds of human error and bias, while at the same time making it more susceptible to other types of mistakes. The primary source of risk, underlying the emergence of PAI, is that it may become misaligned with human interests and goals; given the powers and capabilities of PAI such misalignment can produce catastrophic consequences for humanity. Bostrom noted that we may not be able to predict with precision the moment when PAI will appear.[14]

Critch and Krueger[15] compared the potential effect of PAI to those of the agricultural and industrial revolutions, which have changed the course of humanity. They argued that the emergence of PAI is likely to follow a non-linear path: AI power will grow steadily until one AI system reaches a threshold of self-improvement, at which point it will quickly outperform others by many orders of magnitude, leading to a radical transformation of the current world system.[16] They compared the development of PAI to the spread of humanity on Earth. Just as the non-human world was unable to contend with the expansion of humans, humans will not be able to cope with the powers of PAI.[17] The main risk, Critch and Krueger argued, is that once PAI technology is deployed, it will not be possible to reverse or stop its transformative effect. They noted that whereas machines can operate in a wide spectrum of physical environments, humans can survive only in a relatively narrow set of physical conditions,[18] and are therefore vulnerable to potential changes in the environment that may be beneficial to machines but not to humans.[19]

---

[12] Bostrom Nick, *How long before superintelligence?*, 5 LINGUISTIC AND PHILOSOPHICAL INVESTIGATIONS, 11 (2006).
[13] Bostrom Nick (2014), supra note 2.
[14] Hin-Yan Liu, Kristian Cedervall Lauta & Matthijs Michiel Maas, *Governing Boring Apocalypses: A New Typology of Existential Vulnerabilities and Exposures for Existential Risk Research*, 102 FUTURES 6, 13 (2018); Bostrom Nick, *Ethical Issues in Advanced Artificial Intelligence*, *in* SCIENCE FICTION AND PHILOSOPHY: FROM TIME TRAVEL TO SUPERINTELLIGENCE 277, 284 (2003).
[15] Critch Andrew & Krueger David, supra note 1, at 13.
[16] *Id*, at 18.
[17] *Id*, at 13.

The primary approach of the literature to this risk, which draws on the early writings of Isaac Asimov, is based on the idea of ethical engineering. According to this approach, as AI systems and robots become more advanced and intelligent, they must be subjected to ethical norms.[20] It is useful in this context to distinguish between the *ethics of AI* and *ethical AI*. The ethics of AI deals with the principles that govern the development of AI, ensuring that ethical dilemmas about the interaction of AI with humans and nature are taken into account in the design phase. Ethical AI focuses on the AI system itself: how ethical principles can be incorporated into the AI system, ensuring that it operates ethically in its interaction with other agents (humans, animals, other AI systems).[21]

One of the best-known approaches for *ethical AI* is Isaac Asimov's three laws of robotics, introduced in the 1950s:[22]

1. A robot may not injure a human being or, through inaction, allow a human being to come to harm.
2. A robot must obey orders given it by human beings except where such orders would conflict with the First Law.
3. A robot must protect its own existence as long as such protection does not conflict with the First or Second Law.

We argue that the approach of ethical engineering is problematic in two respects. First, given the condition of moral pluralism it is difficult to see how one can develop an algorithm that can perfectly reconcile between competing moral answers to practical dilemmas. This problem reflects the incommensurability of moral values, where values are considered incommensurate "if it is neither true that one is better than the other nor true that they are of equal value".[23] Advocates of ethical engineering have not given sufficient attention to this problem. Thus, for example, the recent IEEE

[18] *Id*, at 17.
[19] *Id*, at 18.
[20] ISAAC ASIMOV, I, ROBOT (1950); Allen, Colin, Wendell Wallach, and Iva Smit, *Why Machine Ethics?,21 IEEE Intelligent Systems* 12 (2006).

[21] Keng Siau & Weiyu Wang, *Artificial Intelligence (AI) Ethics: Ethics of AI and Ethical AI*, 31 J. OF DATABASE MANAGEMENT, 74 (2020).
[22] Isaac Asimov, supra note 20.
[23] Joseph Raz, *Value Incommensurability: Some Preliminaries, 86* PROCEEDINGS OF THE ARISTOTELIAN SOCIETY 117 (1985).

Standard Model Process for Addressing Ethical Concerns during System Design[24] takes it for granted that values can be successfully prioritized through a process of elicitation and consultation with internal and external stakeholders, and a comparison with external authoritative (regulatory) sources.[25] Human societies have not resolved the problem of incommensurability by developing a universal ranking of values (meta-ethics), but by establishing an intricate network of political, executive and judicial institutions, which through a system of checks and balances, can resolve conflicts between incommensurables.[26]

The second problem with the idea of ethical engineering concerns the risk that a PAI agent may modify its goals while developing its understanding of the world, making them inconsistent with those of its creators.[27] Instilling the AI agent with a system of prioritized values does not provide a robust defense against the risk of misaligned AI creativity.

### B. The intersection between human agency and AI in Facebook internal governance

The activity of Facebook users is subject to a variety of rules set and enforced by Meta. The main set of rules that governs the behavior of Facebook users is the "**Community Standards**."[28] That system is almost fully controlled by Meta, with limited public involvement. Our goal in this section is to expose the intersection between AI tools and human agency in the governance of Facebook's internal platform law. The involvement of algorithms on the Facebook platform extends beyond the enforcement of the community standards (e.g., ad delivery).[29] However, we have limited our

[24] *IEEE Standard Model Process for Addressing Ethical Concerns During System Design: IEEE Standard 7000-2021*, (Sep. 15, 2021), https://ieeexplore.ieee.org/document/9536679.

[25] *Id*, at section 8, 39; Sarah Spiekermann & Till Winkler, *Value-Based Engineering with IEEE 7000*, 41(3) IEEE TECHNOLOGY AND SOCIETY MAGAZINE 71, 75 (2022).

[26] Timothy Endicott, *Proportionality and Incommensurability*, 40/2012 OXFORD LEGAL STUDIES RESEARCH PAPER (2012) Available at SSRN: https://ssrn.com/abstract=2086622.

[27] Wolfhart Totschnig, *Fully Autonomous AI*, 26.5 SCIENCE AND ENGINEERING ETHICS, 2473 (2020); Mitja Kovac, *Autonomous Artificial Intelligence and Uncontemplated Hazards: Towards the Optimal Regulatory Framework*, 13(1) EUROPEAN JOURNAL OF RISK REGULATION, 94 (2022).

[28] Alongside the Community Standards, there are other policies this article will not cover, such as the Advertising Policy that applies to content in commercials, the Page, Groups and Events policy that applies to all creators and managers of pages, groups, and events on Facebook, the Branded Content Policy that applies to business pages, and Recommended Guidelines for Facebook that guide how Facebook promotes content visibility to users. See *Other policies*, META: TRANSPARENCY CENTER (last visited May 27, 2022), https://tinyurl.com/yckwzvhr.

[29] For example, see Ali Muhammad et al., supra note 4; Bobadilla J., Ortega F., Hernando A. & Gutiérrez A., *Recommender systems survey*, 46 KNOWLEDGE-BASED SYSTEMS, 109 (2013); Timothy Garton Ash et al., supra note 4; Fabio Del Vigna, Andrea Cimino, Felice Dell'Orletta, Marinella Petrocchi & and Maurizio Tesconi, *Hate Me, Hate Me Not: Hate Speech Detection on Facebook*, Paper presented at the Proceedings of

analysis to the role of AI tools in the implementation of the community standards because we are using it primarily to illustrate our broader argument about algorithmic constitutionalism. Our analysis relies on publicly available data that probably do not provide a complete picture of the involvement of AI in Facebook compliance processes[30] but this informational gap is not critical for our general argument.

### B.1. Facebook community standards

Facebook community standards govern the behavior of users with respect to content. The standards have been created by Meta's human team, with input from the public and experts in diverse fields such as technology, public safety, and human rights.[31] The purpose of the standards is to create a place for expression and give people a voice on one hand, and on the other to prevent the publication of content that may undermine one of the following objectives: authenticity, safety, privacy, and dignity.[32] Content that violates one of these objectives can still get published if it is newsworthy and in the public interest. Such a determination is made through a formula that balances the damage that may be caused by the publication with public interest in the content.[33]

The standards are divided into six main sections dealing with: violence and criminal behavior, safety, objectionable content, integrity and authenticity, respecting intellectual property, and content-related requests and decisions (see Table 1 below for details). For each section, the standards elaborate the differences between various categories of content according to the following schema:[34]

- Content that is not allowed
- Content that requires additional information or context, content that is allowed with a warning screen, or content that is allowed but can be viewed only by adults aged 18 and older the First Italian Conference on Cybersecurity (2017), https://ceur-ws.org/Vol-1816/paper-09.pdf; Huang S., Zhang J., Wang L. & Hua X, *Social Friend Recommendation Based on Multiple Network Correlation*, 18(2) IEEE TRANSACTIONS ON MULTIMEDIA, 287 (2016); Sean MacAvaney et al., supra note 4.

[30] For example, in a study conducted by Kate Klonick, significant gaps were discovered between the declared community standards and the rules according to which the Facebook review teams intervened in content. See Kate Klonick, *The New Governors: The People, Rules, and Processes Governing Online Speech*, 131 HARV. L. REV., 1598, 1619 (2018).

[31] See *Facebook Community Standards*, META: TRANSPARENCY CENTER (last visited May 27, 2022), https://tinyurl.com/exby8z7p.

[32] *Id*.

[33] See *Our Approach to Newsworthy Content*, META: TRANSPARENCY CENTER, (Jan. 19, 2022), https://tinyurl.com/3by86296.

[34] See Facebook Community Standards, supra note 31.

For example, in the section on Violence and Incitement, content that users cannot post and that will be removed includes: "Statements of intent to commit high-severity violence;" "Calls for high-severity violence;" "Aspirational or conditional statements to commit high-severity violence." Some threatening language, however, requires additional information or context for Meta to act upon, such as "Violent threats against law enforcement officers" or "Violent threats against people accused of a crime."[35] Table 1 summarizes the content categories governed by the community standards.

**Table 1: Content categories covered by the community standards**

| **Violence and criminal behavior** | **Safety** | **Objectionable content** | **Integrity and authenticity** | **Content-related requests and decisions** | **Respecting intellectual property** |
|---|---|---|---|---|---|
| Violence and incitement | Suicide and self-injury | Hate speech | Account integrity and authentic identity | User requests | Intellectual property |
| Dangerous individuals and organizations | Child sexual exploitation, abuse, and nudity | Violent and graphic content | Spam | Additional protection of minors | |
| Coordinating harm and promoting crime | Adult sexual exploitation | Adult nudity and sexual activity | Cybersecurity | | |
| Restricted goods and services | Bullying and harassment | Sexual solicitation | Inauthentic behavior | | |
| Fraud and deception | Human exploitation | | Misinformation | | |
| | Privacy violations | | Memorialization | | |

### B.2. The hybrid (AI/human) Facebook internal compliance system

The community standards are enforced by a hybrid compliance system that combines AI algorithms and human agents (users' input, review teams, and external experts). The enforcement process consists of four stages (**Figure 1**):

1) Preliminary (*prima facie*) flagging of content as breaching the community standards.

[35] See *Violence and Incitement*, META: TRANSPARENCY CENTER, (last visited May 27, 2022), https://tinyurl.com/2p9c4bvp.

2) Investigation of the preliminary flagging and a determination whether a violation has in fact occurred.

3) Imposition of sanction or other administrative response.

4) Re-examination and appeal concerning the punitive action.

In addition to these four steps Meta has employed in the past few years a system called cross-check, which provides some categories of entities and content special consideration and is governed by a distinct set of rules. Below we briefly describe how AI and human agents intersect in the enforcement process.

**1) Preliminary (*prima facie*) finding of a violating content**

The enforcement process is initiated by flagging an item of content as potentially violating. Such flagging can occur either through user reports or through AI detection, based on NLP technology.Every post on Facebook (and Instagram) is scanned [36] immediately after its publication.[37] To enhance its capacity to detect violating content, Meta developed an AI model called "Few-Shot Learner" (**FSL**). Traditional AI models require a massive amount of labeled data points to train the model. Collection and labeling of these data can take several months.[38] Under FSL, violating content can be detected even without the availability of a large body of labeled data. FSL can respond to new types of harmful content within weeks instead of months. It can work in more than 100 languages and is able to learn from various types of data, including images and text.[39] Meta describes this process as follows: "First, it is trained on billions of generic and open-source language examples. Then, the AI system is trained with policy-violating content and content that were labeled over the years. Finally, it's trained on condensed text explaining a new policy… the key idea is to convert the class label into a natural language sentence which can be used to describe the label, and determine if the example entails the label description."[40]

[36] See *How Meta Enforces its Policies*, META: TRANSPARENCY CENTER (last visited May 27, 2022), https://tinyurl.com/yjf27rek; see further: Goldstein, I., Edelson, L., McCoy, D., & Lauinger, T., *Understanding the (In) Effectiveness of Content Moderation: A Case Study of Facebook in the Context of the US Capitol Riot. arXiv preprint arXiv:2301.02737* (2023).
[37] See *How Meta Prioritizes Content for Review*, META: TRANSPARENCY CENTER (last visited May 8, 2022), https://tinyurl.com/yckmhwf2.
[38] See *Harmful Content Can Evolve Quickly. Our New AI System Adapts to Tackle it*, META AI (last visited May 27, 2022), https://tinyurl.com/fm2388j3.
[39] *Id*.
[40] *Id*.

FSL can also anticipate new types of policy violations. This information can allow Meta to "understand the sensibility and validity of the proposed definitions. It casts a wider net, surfacing more types of 'almost' content violations that policy teams should know about when deciding or formulating at-scale guidance for annotators who train new classifiers and the human reviewers helping to keep our platform safe… Our long-term vision is to achieve human-like learning flexibility and efficiency, making its integrity systems even faster and easier to train and better able to work with new information."[41]

**2) Investigation of a *prima facie* flagging and determination of violation**

After content has been flagged as potentially in violation, a determination must be made whether it violates the community standards. This determination can be made either by AI algorithm or a review team. Usually, content flagged as potentially violating (by AI or by user reports) is submitted for review by the AI tool to determine whether it is indeed violating. There are five potential responses:

1) There is a violation.

2) There is no violation.

3) There is no violation, but the content is sensitive or misleading.

4) There is no violation, but the content is problematic.

5) Unable to decide if there is a violation.

If it is determined that there was no violation, the suspicious content is 'acquitted' and the inquiry is concluded. If the content is found to be violating a decision is required regarding the appropriate sanction. If the content is determined to be sensitive or misleading (but not violating), it will usually not be removed, and less severe actions aimed at protecting other users will be taken. Content is classified as problematic when it does not violate community standards but may nevertheless harm users and impair their experience of using Facebook,[42] for example, when the content contains links to sites that are packed with advertisements, or when it leads to shocking, disturbing, or malicious advertisements.[43] The same is true of content that constitutes a "clickbait,"[44] that is, content that includes misleading, sensational, or spammy stories whose headlines are intended to grab users' attention in order to click on a link.

[41] *Id*.
[42] See *Reducing the Distribution of Problematic Content*, META: TRANSPARENCY CENTER (last visited May. 24, 2022), https://tinyurl.com/5cur2xwu.

When the AI algorithm fails to decide whether there has been a violation, the content is passed on to human review teams for determination. This happens in borderline cases, where the wording is vague or the context is important for reaching a decision.[45] The review teams consist of trained human examiners, who have expertise in the domain of the suspected violation and in the language in which the content was published. The review teams intervene in the decision-making process in the following cases: (a) when the AI algorithm has been unable to determine whether there is a violation; (b) when the suspicious content has been flagged by users; (c) in content categories that were predefined by Meta as necessitating human review, e.g., content that requires attention to subtle nuances (such as bullying);[46] (d) content that falls under Meta's cross-check system (see below for extended elaboration). The order through which the review teams consider the flagged posts is determined by AI. The AI algorithm considers three factors in deciding how to prioritize the flagged content: (a) the likelihood that it violates community standards; (b) the probability that its publication will cause harm; and (c) its virality.[47]

The review teams can choose between the five responses described above. If the review team is unable to decide whether a violation has occurred, it can consult with specialized experts from the Global Operations or Content Policy Teams.[48] Meta notes that it uses the information obtained from the decisions of the human review teams to train and improve the accuracy of its AI algorithm.[49]

*The Cross-Check System*

The existence of the cross-check system was exposed by the Wall Street Journal on September 2021. The report, which was based on documentation produced by former employee and company critic Frances Haugen, described the system as exempting Meta's most influential users from normal content moderation processes.[50] The Oversight Board received a request from Meta to review the system and consider how it could be improved.[51] The following analysis is based on the Board advisory opinion which was published on 6 December 2022.[52] Meta has not been transparent about the system, and its details have become public only after the publication of the board's recommendations.[53] The cross-check system provides additional layers of review for two types of predefined content in an effort to prevent over-enforcement or false-positives. First, cross-check provides guaranteed additional human review of content that is created by specific entities.[54] Under this scheme, which is called **Early Response Secondary Review (ERSR),** if a protected entity posts content that is identified as violating, it will not be removed according to the procedures that apply to regular users, but instead will be sent for extra levels of review. The second scheme that forms part of the cross-check system is **General Secondary Review** (**GSR**). Whereas ERSR applies to all content posted by specific entitled entities, GSR may apply to any content posted on the platform, based on a determination by the cross-check ranker algorithm, that it merits additional review because of a high likelihood of being a false positive. The cross-check ranker algorithm relies on the following features: "topic sensitivity (how trending/sensitive the topic is), enforcement severity (the severity of the potential enforcement action), false positive probability, predicted reach, and entity sensitivity".[55]

---

[43] See *Reducing Links to Low-Quality Web Page Experiences*, META: TRANSPARENCY CENTER (last visited May 10, 2017), https://tinyurl.com/djhrx3wf.
[44] See *New Updates to Reduce Clickbait Headlines*, META: TRANSPARENCY CENTER (last visited May 17, 2017), https://tinyurl.com/5782un2w.
[45] See How Meta Prioritizes Content for Review, supra note 37.
[46] See *How Review Teams Work*, META: TRANSPARENCY CENTER (last visited Jan. 19, 2022), https://tinyurl.com/24bhmp8a.
[47] See How Meta Prioritizes Content for Review, supra note 37.
[48] See How Review Teams Work, supra note 47.
[49] See How Meta Prioritizes Content for Review, supra note 37.

In both categories, the cross-check process kicks in after a piece of content is flagged for review by AI or is reported by a human user, and before the imposition of any sanction.[56] The additional review is performed by human moderators.[57] One of the issues that was criticized by the board was the unequal treatment received by content in the ERSR and GSR streams: whereas content that is reviewed under the ERSR can benefit from up to four review cycles, content that is reviewed under the GSR can benefit from maximum two review cycles.[58] The enforcement actions against content recognized as eligible for cross-check will be suspended until the end of the review process. Under the current cross-check system, content reviewed under ERSR receives priority over GSR content. Due to the limited capacity of the Early Response Teams, not all GSR content is reviewed.[59]

---

[50] Decision, p. 6.
[51] Decision, p. 3.
[52] Oversight Board, *Policy advisory opinion on Meta's cross-check program*, 25-35 (December 2022), https://www.oversightboard.com/decision/PAO-NR730OFI/. See also, see Kate Klonick, *The Meta Oversight Board Has Some Genuinely Smart Suggestions*, THE ATLANTIC (Dec. 9, 2022), https://tinyurl.com/54sftprj.
[53] See Policy advisory opinion on Meta's cross-check program, supra note 52, at 3.
[54] *Id*, at 9-15. Meta reported to the Oversight Board that it includes entities on ERSR lists according to the following categories: "(1) civic and government; (2) significant world events; (3) media organizations, businesses, communities and creators, including advertisers; (4) historically over-enforced; (5) legal and regulatory or entities for which erroneous action may present legal risk to Meta, for example in the context of ongoing litigation; (6) entities whose content is under review, meaning cases where action by any reviewer would undermine ongoing deliberation or would present risk to Meta". *Id*, at 10.
[55] *Id*, at 19.
[56] *Id*, at 14, 21.
[57] *Id*.

The Oversight Board exposed another system that blocks some enforcement actions outside of the cross-check system, which Meta referred to as "technical corrections". "Technical corrections" are automatic exceptions to content policy enforcement. As explained by Meta, a "technical correction" applies only to a specific entity for a specific policy violation and does not serve to bar enforcement for other policy violations. At the time of the Board's briefings with Meta, it stated that it applied about a thousand technical corrections per day. Meta did not disclose how many and what type of entities have benefited from a "technical correction".[60]

The Oversight Board has strongly criticized the cross-check program, highlighting its unequal nature (which was manifested in the privileged treatment of business partners and the lack of proper protection to entities that are likely to produce expressions with significant human rights value) and its non-transparent functioning (e.g., Meta's refusal to share data regarding the identity of the business partners enjoying cross-check privileges).[61]

**3) Compliance action**

Once a decision has been made whether an item of content violates the community standards, compliance action is taken against the user. A decision on that action can be made by either a review team or by AI algorithm.[62] The nature of the sanction or other action will be determined based on the circumstances and by the nature of the "offense." There is a broad spectrum of potential compliance responses with varying levels of severity:

- complete removal of the infringing content;
- Limiting the exposure of the content by reducing its News Feed ranking (e.g., in the context of false information).
- Restricting the visibility of content for some users, for example, those under 18 (e.g., in the case of content related alcohol, tobacco, sex toys, or weapons).
- Adding a warning screen/tag (e.g., in relation to content that refers to non-consensual sex).[63]
- Disabling the account in cases of serious violations, such as the publication of content that contains child pornography or physical abuse.[64]
- Reporting to relevant law-enforcement authorities.[65]

[58] *Id*, at 14, 21.
[59] *Id*, at 19-20. Unlike ERSR, content in the GSR is put in a queue pending review. Content that is not reviewed within 2-4 days is 'popped out' of the GSR queue. When this happens, Meta reverts to its initial enforcement decision (e.g., removal or a warning screen) without further review. *Id*, at 20.
[60] *Id*, at 22.
[61] *Id*. The Board provided Meta with a series of detailed recommendations, *id*, at 35-48.
[62] Meta states that it uses a dedicated enforcement technology system that determines whether to take action, but does not specify in which cases decisions are made by algorithms and in which they are made by human teams. See *How enforcement technology works*, META: TRANSPARENCY CENTER (last visited Nov. 14, 2022), https://tinyurl.com/mvw2edf4 and Díaz Á. & Hecht-Felella L., *Double standards in social media content moderation*, BRENNAN CENTER FOR JUSTICE AT NEW YORK UNIVERSITY SCHOOL OF LAW (2021),

When users are involved in repeated violations, they may be subject to additional sanctions, such as restricting accounts for a limited period or in severe cases, disabling the account.[66] Meta uses the method of counting strikes in determining the severity of the measures it takes against users. The system keeps a record of the violations (strikes) of each user. If the number of strikes exceeds a certain threshold, it may lead to the restriction or disabling of the account,[67] which means that users' ability to create content (e.g., posting, commenting, creating a Page ) would be subject to limitations.[68]

The first strike usually results in a warning, with no further action. Two strikes may lead to restrictions limited in time (e.g., one day), three strikes to three days, and four strikes to seven days. From the fifth strike onward, the account may be limited

https://www.brennancenter.org/our-work/research-reports/double-standards-social-media-content-moderation.

[63] See *Adult Sexual Exploitation*, META: TRANSPARENCY CENTER (last visited May 27, 2022), https://tinyurl.com/2p8rwufv.

[64] See *Child Endangerment: Nudity and Physical Abuse and Sexual Exploitation*, META: TRANSPARENCY CENTER (last visited May 27, 2022), https://tinyurl.com/mt7aj89f.

[65] *Id.*

[66] See *Restricting Accounts by Public Figures During Civil Unrest*, META: TRANSPARENCY CENTER (last visited May 26, 2022), https://tinyurl.com/2p9y995d.

[67] See *Counting Strikes*, META: TRANSPARENCY CENTER (last visited May 25, 2022), https://tinyurl.com/3ftwvkec.

[68] See *Restricting Accounts*, META: TRANSPARENCY CENTER (last visited May 26, 2022), https://tinyurl.com/aa7nymbc.

for 30 days for each further strike,[69] and may be entirely disabled, depending on the severity of the violation.[70] There may be cases of serious violations in which an account is disabled after one strike, as in the case of posting child pornography. There are certain types of accounts that Meta disables as soon as it notices their existence, such as accounts of convicted sex offenders or accounts that display false identities.[71]

The rules applicable to groups and to pages that consistently violate community standards differ from those that apply to private users.[72] For example, Meta may require page and group managers to approve each post before it is published.[73] During civil unrest, Meta may restrict accounts of public figures for longer periods, if it finds that they incite violence.[74]

Meta notifies users when they post content that violates community standards. The notice of infringement usually includes a mention of the standard that was violated and a brief description of the reason why the content was removed.[75] When an account is disabled, Meta notifies the user when trying to log into the account.[76] Meta also notifies the user when it issues a "strike" and when the account has been restricted.[77]

#### 4) Appeal and reexamination procedures

If users believe that they have been subject to a compliance action for no justified reason, and that the content that they shared does not violate Facebook's community standards, they can request that the post be reexamined.[78] The reexamination process usually takes up to 24 hours.[79] If the reexamination reveals that there was a mistake and the content was removed or concealed without a valid reason, the content is restored and the user is notified.[80] The review is conducted by human audit teams.[81]

---

[69] *Id*.
[70] See *Disabling Accounts*, META: TRANSPARENCY CENTER (last visited May 26, 2022), https://tinyurl.com/3ueruprd.
[71] *Id*.
[72] For more information, see *Removing Pages and Groups*, META: TRANSPARENCY CENTER (last visited May 26, 2022), https://tinyurl.com/5yp54tbm.
[73] See Restricting Accounts, supra note 68.
[74] See Restricting Accounts by Public Figures During Civil Unrest, supra note 66.
[75] See *Taking Down Violating Content*, META: TRANSPARENCY CENTER (last visited May 26, 2022), https://tinyurl.com/2p897w74.
[76] See Disabling Accounts, supra note 70.
[77] See Restricting Accounts, supra note 68.
[78] See *I Don't Think Facebook Should Have Taken Down my Post*, FACEBOOK: HELP CENTER (last visited May 27, 2022), https://tinyurl.com/nrkfj85z. Facebook does not specify in which cases the right to have the action reexamined is not granted.
[79] *Id*.

Users who do not agree with the outcome of the reexamination process can appeal to the Oversight Board within 15 days from the day on which they received the final decision, which reviews decisions by Facebook and Instagram as well as cases referred to it by Meta.[82] The Oversight Board has the right to choose which cases it reviews.[83] Its decisions are binding, and Meta is committed to implementing them.[84] The Board is overseen by a group of trustees.

According to its bylaws, the Oversight Board must set up two subcommittees, the Membership Committee and the Case Selection Committee, and may establish additional ones. The case selection committee will set criteria for the cases that the board will prioritize and select for review. Its decisions will be by a majority vote, subject to override by a majority vote of the full board.[85] The Oversight Board can review only a very small portion of the total number of Facebook content moderation decisions. The board will aim to issue its decisions within ninety days.[86] Each case is preliminarily reviewed by a panel of five members that can request more information from Facebook and other sources, such as external experts or advocacy groups that reflect a range of perspectives. The panel receives a statement from the user who submitted the request; the case history (delivered by Meta); relevant policy (also from Meta); and any additional external information that any member of the panel may ask for.[87]

Meta is committed to applying Board decisions beyond the case being reexamined. For example, if the Board decides that a certain content violates the community standards, Meta will conduct a search aimed at identifying similar content, and if it is technically feasible, will take measures against it.[88]

---

[80] *Id*.
[81] See *Appealed Content*, META: TRANSPARENCY CENTER (last visited May 27, 2022), Appealed content | Transparency Center (fb.com).
[82] *Oversight Board Bylaws*, February 2023, Article 3, 1.1 , https://www.oversightboard.com/sr/governance/bylaws.
[83] *Id,* Article 1, section 3.1.2 (Selection of Cases)..
[84] Except when the implementation of Board decisions would be against the law. *Id* Article 2, section 2.3.
[85] *Id*, Article 1, section 1.2.1.
[86] *Id*, Article 1, section 3.1.2.
[87] *Id*, Article 1, section 3.1.5.
[88] *Id*, Article 2, section 2.3.1.

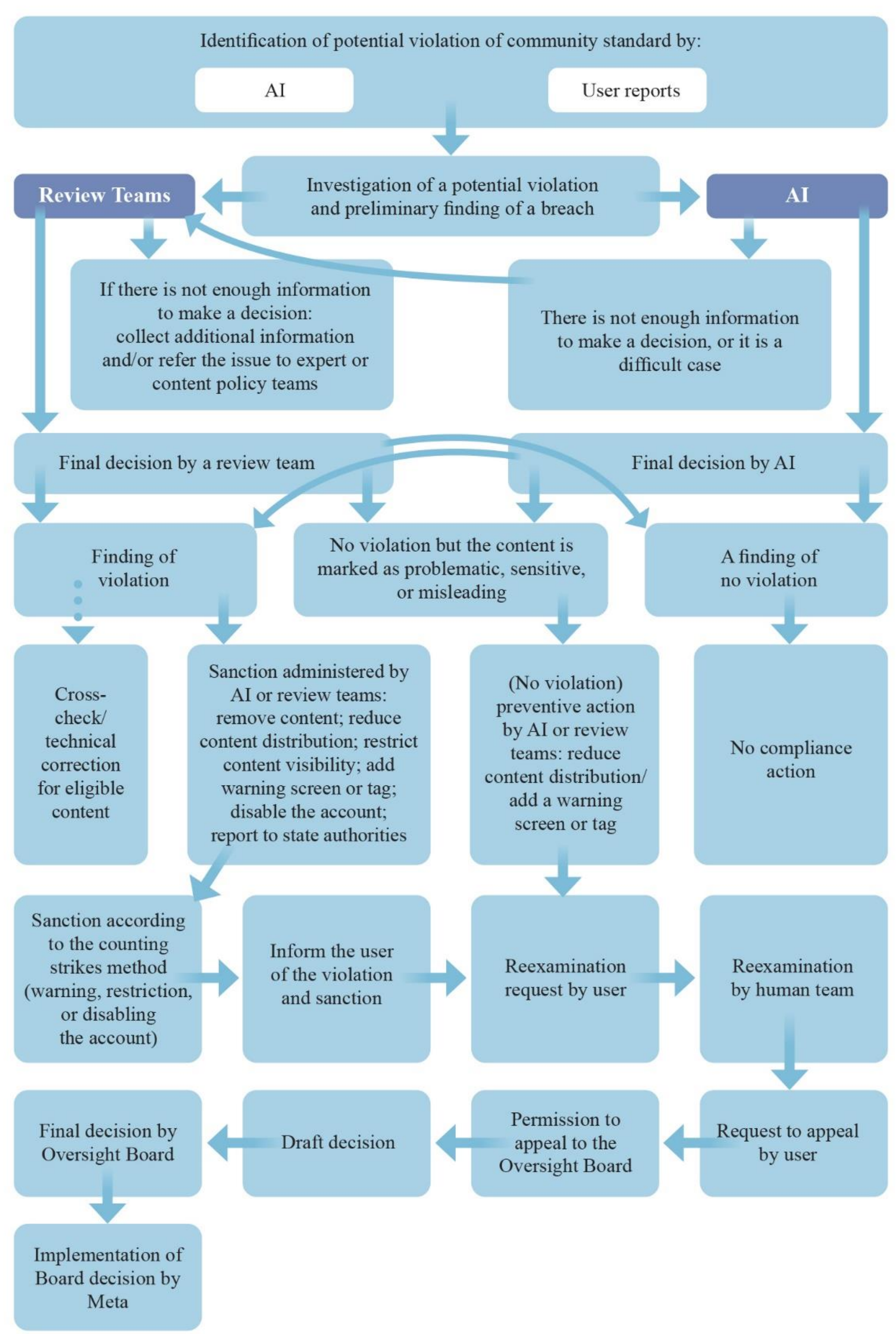

**Figure 1: Description of Facebook's content moderation framework**

## C. The risks of a shift to full AI governance: An imaginary Facebook with no humans in the loop

Facebook's compliance framework is a hybrid system in which human agents interact with AI algorithms. In this section, we explore the potential risks of an imaginary scenario in which Facebook becomes governed entirely by AI and humans cease to be part of the compliance loop. We assume that this AI system possesses part of the qualities of a PAI, in particular, a capacity to make autonomous decisions that may be misaligned with the objectives of its controllers. A shift to full AI governance raises two types of risks: potential increase in cases of over and under enforcement (false positives and false negatives) due to loss of human oversight and algorithmic bias, and potential misaligned, AI-driven revision of the system's fundamental normative structure. This hypothetical scenario seems like a natural extension of existing practices on social media platforms, which already use algorithms extensively to manage their operations. This reliance allows them to scale up their operations and cut costs

One of the key implications of removing humans from the compliance loop is the loss of the distinction between preliminary, first-order, compliance actions and second-order, review/appeal decisions. Most of the preliminary compliance decisions at Facebook are made by automated means. Human involvement in the preliminary stage is limited to (a) difficult or borderline cases and (b) failure of the AI algorithm to decide whether a certain content violates the standards.[89] The second-order review stage is conducted entirely by humans. If human agents are no longer involved in the compliance process, the distinction between the two stages of decision will disappear, resulting in an integrated automated compliance process.

The question is whether a shift to a fully automated compliance process would lead to increase in the number of false positives and false negatives, in a way which adversely affect users' welfare and rights. In their recent book, *Noise: A flaw in human judgment,* Kahneman, Sibony, and Sunstein argued that algorithmic prediction techniques represent "significant improvements on human judgment. The combination of personal patterns and occasion noise weighs so heavily on the quality of human judgment that simplicity and noiselessness are sizable advantages."[90] As Jhaver and colleagues noted in a study about content moderation on Reddit, "[a]utomating moderation not only facilitates scalability, it also enables consistency in moderation decisions."[91]

[89] See page 12 above.

Other works, however, have offered an opposing perspective, which emphasizes the risk of algorithmic bias, which occurs when "the outputs of an algorithm benefit or disadvantage certain individuals or groups more than others without a justified reason for such unequal impacts."[92] Such bias can have several causes: faulty design, a training dataset that is contaminated with social biases (either owing to faulty sampling or because of biases entrenched in society), or bias-reinforcing loops.[93] Researchers and activists have pointed out algorithmic bias in a variety of environments, including criminal justice, healthcare, hiring, and credit scores.[94]

Meta has been involved in several cases of algorithmic bias, which demonstrate that the foregoing concerns are not unfounded. Multiple studies have shown that the Facebook ad delivery system can be skewed by gender or race because of hidden algorithmic optimization, even when not requested by the advertisers.[95] A recent lawsuit filed against Meta by the US Justice Department alleged that housing advertisements on Meta Platforms have discriminated users based on their race, color, religion, sex, disability, familial status and national origin, in violation of the Fair Housing Act. As part of a settlement agreement signed with Meta in June 2022, Meta agreed to stop using the advertising tool known as the "Special Ad Audience" that had a discriminatory effect. Meta has also agreed to develop a new system to address racial and other disparities caused by its use of personalization algorithms in its ad delivery system for housing ads.[96] A recent study published by the NYU Brennan Center for Justice has shown that the Facebook content moderation approach to hate speech and violent extremism content was biased against minority groups.[97]

[90] KAHNEMAN DANIEL, OLIVIER SIBONY & CASS R. SUNSTEIN, NOISE: A FLAW IN HUMAN JUDGMENT, 133 (2021). We must distinguish between noise, which "consists of unwanted variability in judgments," and bias, which "consists of any systematic error that inclines people's judgments in a particular direction," Sunstein C. R., *Governing by Algorithm? No Noise and (Potentially) Less Bias*, 71 DUKE LAW JOURNAL, 1175, 1178 (2022).

[91] Jhaver, S., Birman, I., Gilbert, E., & Bruckman, A., *Human-Machine Collaboration for Content Regulation: The Case of Reddit Automoderator*, 26(5) ACM TRANSACTIONS ON COMPUTER-HUMAN INTERACTION (TOCHI) 1 (2019).

[92] Kordzadeh Nima, and Maryam Ghasemaghaei, *Algorithmic Bias: Review, Synthesis, and Future Research Directions*, 31 no. 3 EUROPEAN JOURNAL OF INFORMATION SYSTEMS, 388, 388 (2022).

[93] *Id*, at 390; Baker Ryan S. & Aaron Hawn, *Algorithmic Bias in Education*, INTERNATIONAL JOURNAL OF ARTIFICIAL INTELLIGENCE IN EDUCATION, 1, 2 (2021); CHRISTIAN BRIAN, THE ALIGNMENT PROBLEM: MACHINE LEARNING AND HUMAN VALUES, 33-34 (2020).

[94] *Id*.

[95] Basileal Imana, Aleksandra Korolova, and John Heidemann. 2021. Auditing for Discrimination in Algorithms Delivering Job Ads. In Proceedings of the Web Conference 2021 (WWW '21), April 19–23, 2021, Ljubljana, Slovenia. ACM, New York, NY, USA, 3767; Kaplan, L., Gerzon, N., Mislove, A., & Sapiezynski, P. (2022, October). Measurement and analysis of implied identity in ad delivery optimization. In *Proceedings of the 22nd ACM Internet Measurement Conference* (pp. 195-209).

The problem with removing human agency from the Facebook compliance system is that there is no general algorithmic de-biasing tool that can reliably replace human oversight. This reflects the complexity of algorithmic bias, its diverse manifestations, and multiple potential causes.[98] Making the Facebook compliance system more consistent and noiseless runs the risk of uncontrolled algorithmic bias.

The second type of risk that may result from entrusting the governance of Facebook to an AI system emanates from its semi-prepotent capacities: its ability to initiate, on its own "volition," a process leading to the revision of Facebook's fundamental norms in a direction that may be incompatible with fundamental human rights. This risk has several aspects that should be distinguished. The first aspect focuses on the potential adverse effects of an overzealous, unreflective optimizing AI agent. Such an agent could pursue its own interpretation of an underspecified objective or rule and the means for realizing it, in ways and scales that were not foreseen by the designer. The second aspect concerns the risk that an autonomous AI agent may develop novel objectives that are distinct (and potentially misaligned) from those that were given to it by its designer.[99]

We can illustrate these risks by considering the Facebook mission statement, "to give people the power to build community and bring the world closer together."[100] Facebook has changed its mission statement several times since it was established.[101]

---

[96] United States of America v. Meta Platforms Inc, 22 Civ. 5187, Settlement Agreement (21.06.2022), https://tinyurl.com/5n6nmtkf.

[97] Díaz et al., supra note 62, at 5-6, 10-11.

[98] Raghavan Manish, Solon Barocas, Jon Kleinberg & Karen Levy, *Mitigating Bias in Algorithmic Hiring: Evaluating Claims and Practices*, *In* PROCEEDINGS OF THE 2020 CONFERENCE ON FAIRNESS, ACCOUNTABILITY, AND TRANSPARENCY, 469, 477F (2020); Bent Jason R., *Is Algorithmic Affirmative Action Legal*, 108 GEO. LJ 803, 814-816 (2019).

[99] Ecoffet Adrien, Jeff Clune & Joel Lehman, *Open Questions in Creating Safe Open-Ended AI: Tensions Between Control and Creativity*, *in* ALIFE 2020: THE 2020 CONFERENCE ON ARTIFICIAL LIFE, 27 (2020); Bostrom Nick (2003), supra note 14. In a recent interesting paper, Muller and Cannon argue that it is unplausible to attribute the risk of unreflective optimization to PAI with general intelligence, since a capacity for self-reflection is a key feature of such intelligence. What we should worry about, they argue, is the (significant) damage that can be caused by a highly capable instrumental AI (not a PAI), if designed or used badly. See: Müller, Vincent C., and Michael Cannon. "Existential risk from AI and orthogonality: Can we have it both ways?." 35 *Ratio* (2022): 25, 34.

[100] See *FAQs*, META INVESTOR RELATIONS (last visited Sep. 17, 2022) https://tinyurl.com/32fh6d87.

[101] The previous version of the mission statement was: "*Facebook gives people the power to share and make the world more open and connected*". Nicholas A. John, *Social Media Bullshit: What We Don't Know About*

The last change was made in 2017.[102] The Facebook mission statement uses broad and vague language, and it is easy to imagine how an attempt to optimize them can generate results that would be considered morally problematic. The mission statement does not specify the types of communities it seeks to support or the means that can be used to bring people together. How will our imagined AI controller respond to an emerging community of neo-Nazis or ISIS supporters? What strategy will it adopt if it finds that fake news regarding climate change or vaccines can be effective in bringing people together? These are not just hypothetical scenarios, as Facebook has already been implicated in facilitating the dissemination of fake news regarding the COVID-19 epidemic and climate change.[103] A recent study by the Institute for Strategic Dialogue has found that Holocaust denial content has been actively recommended by the Facebook algorithm, and that when a user follows public pages containing Holocaust denial content, Facebook actively promotes further Holocaust denial content to that user.[104] Our optimizing AI controller may even go further in seeking to expand Facebook communities by abolishing rules that in its view constitute an impediment to such expansion, e.g., the rule that prohibits hate speech or the policy that reduces the visibility of false news.[105] The problem may become even more acute if the AI agent revises Facebook overarching mission in a way that further supports such perverse expansion of communitarian communication.

### D. Algorithmic Constitutionalism

In this section we introduce and develop the concept of algorithmic constitutionalism and elaborate the ways in which it can be implemented on a social media platform such as Facebook. The argument that follows is highly speculative, leaving many

*Facebook.com/peace and Why We Should Care*, 5.1 SOCIAL MEDIA + SOCIETY, 13 (2019), https://journals.sagepub.com/doi/pdf/10.1177/2056305119829863.

[102] In June 2017, Mark Zuckerberg published a post on his personal Facebook page, in which he announced the decision to change Facebook's mission statement and the reasons that led to this change. See Mark Zuckerberg's official Facebook account. See *Bringing the World Closer Together*, (published on Jun. 22, 2017, last edited on Mar. 15, 2021, last visited Oct. 17, 2022), https://www.facebook.com/notes/393134628500376/.

[103] Edelson, Laura, Minh-Kha Nguyen, Ian Goldstein, Oana Goga, Damon McCoy, and Tobias Lauinger, *Understanding engagement with US (mis) information news sources on Facebook*, In PROCEEDINGS OF THE 21ST ACM INTERNET MEASUREMENT CONFERENCE, 444 (2021).

.

[104] Jakob Guhl & Jacob Davey, *Hosting the 'Holohoax': A Snapshot of Holocaust Denial Across Social Media*, Institute for Strategic Dialogue (2020), https://www.isdglobal.org/wp-content/uploads/2020/08/Hosting-the-Holohoax.pdf

[105] See *Hate Speech and False News*, META: TRANSPARENCY CENTER (last visited March 21, 2023); https://transparency.fb.com/policies/community-standards/hate-speech/; https://transparency.fb.com/policies/community-standards/false-news/. .

technical questions open. Much of the technological details underlying our argument are still at an early stage of development. The concept of algorithmic constitutionalism is based on three pillars: (a) **layered architecture**, whose goal is to insulate the core principles of the system, located at the meta-code level, from self-initiated changes that can modify the fundamental features of the system. We conceive the system as being comprised of two levels of code: (i) operative or object level and (ii) meta level; (b) **algorithmic meta-reasoning**, which conceptualizes the system as operating simultaneously at two levels, allowing it to self-monitor, verify, and potentially correct in real-time operations at the object level if they depart from the principles protected by the meta-level code. We figuratively described the meta-level code as a restraining bolt;[106] and (c) **correction by deliberation**, which would limit the ability of the meta-level code to initiate corrective decisions that affect the fundamental features of the system, and resort instead to deliberation as a mechanism for making such decisions. It is at this junction that societal and algorithmic constitutionalism can collide. Algorithmic constitutionalism thus establishes a hybrid framework, in which human agents and algorithms collaborate, in a way which is critical for the platform's safety, legitimacy, and robustness.[107]

The concept of algorithmic constitutionalism has some affinities with Stuart Russell's principles of beneficial AI, which are:[108]

1. The only objective of the machine is to maximize the realization of human preferences.
2. The machine is initially uncertain about what those preferences are.
3. The ultimate source of information about human preferences is human behavior.

Consistent with Russell's argument, we want to instill in the AI governance system uncertainty about human preferences, which consequentially would be reflected in uncertainty about the correct or optimal interpretation of the rules (in our case, Facebook community standards) and of the overarching mission of the system. As Russell has argued, we want to design a *humble* AI that is willing to defer to humans:[109]

---

[106] De Giacomo et al., supra note 9; Costantini Stefania, supra note 8.
[107] Jhaver et al., supra note 91, similarly emphasize the advantages of mixed-initiative regulation systems where humans work alongside automated systems (at 28).
[108] RUSSELL STUART, HUMAN COMPATIBLE: ARTIFICIAL INTELLIGENCE AND THE PROBLEM OF CONTROL, Ch. 7 (2019).
[109] *Id*.

> A machine that assumes it knows the true objective perfectly will pursue it single-mindedly. It will never ask whether some course of action is OK, because it already knows it's an optimal solution for the objective. It will ignore humans jumping up and down screaming, "Stop, you're going to destroy the world!" because those are just words. Assuming perfect knowledge of the objective decouples the machine from the human: what the human does no longer matters, because the machine knows the goal and pursues it. On the other hand, a machine that is uncertain about the true objective will exhibit a kind of humility: it will, for example, defer to humans and allow itself to be switched off.

In contrast to Russell, we reject the idea that the ultimate source of information about human preferences should be human behavior. Rather, we argue that the system should have a variety of procedures *embedded* in its algorithmic structure by which it can receive input from human agents (users) about the application of the standards, the system's fundamental characteristics, and potential revisions of rules.

Our approach also differs from that of Russell in that we reject the idea that the AI system should seek to optimize some measure of human preferences. Optimization coupled with super-intelligence is a recipe for trouble, as Russell, Bostrom, and others have demonstrated.[110] The risk is most severe when relentless optimization comes with a limited capacity for self-reflection and incomplete rules. Optimization was never adopted as an overarching principle for the governance of human societies. Democratic societies are governed by the somewhat arbitrary rule of politics, which ultimately settles disagreement by a binary decision. This inherent arbitrariness is countered by a system of checks and balances that guarantees the reflexivity of the system and prevents it from becoming captured by a single voice, and in which independent judiciary plays a crucial role.[111]

Instead of optimizing, we propose the idea of *political satisficing*. The idea of satisficing was introduced by Herbert Simon who argued that individuals do not choose alternatives as if they were maximizing (e.g., a utility function) but rather act as satisficers: decision makers who consider a limited number of alternatives, investing limited cognitive effort, until they find one that is "good enough," at which point they stop searching. A *politically satisficing AI agent* is one whose objective is not to maximize human preferences but rather to satisfice them subject to deliberative

[110] Gans Joshua S., *Self-Regulating Artificial General Intelligence*, w24352 NATIONAL BUREAU OF ECONOMIC RESEARCH (2018).
[111] Huysmans Jef, *International Politics of Exception: Competing Visions of International Political Order Between Law and Politics*, 31.2 ALTERNATIVES 135 (2006).

constraints, which are used to identify human preferences and resolve ambiguities about them.

The challenge is to elaborate how this abstract framework can be implemented in practice. The first step is to distinguish between the operative (base) level, which covers the implementation of Facebook community standards (or those of a similar social media platform), and the meta level. The object level AI agent has a broad leeway for managing and improving the application of the standards but is not permitted to explicitly change their wording or revise the procedural principles they establish (e.g., right to appeal). Thus, the object level algorithm is free to try to improve its classification of hate speech and violent extremism content, for example, by reducing the percentage of false positives (i.e., cases wrongly classified as violating the standards) and by improving the sensitivity of the system to social contexts (e.g., to the unique discourse of minority communities).[112] But the object level AI cannot revise the rules themselves (e.g., abolish the policy dealing with hate speech).

The meta-level code acts as a guardian of the community standards, which encode the core values of the community. The meta-level code also governs the process through which the standards may be revised. At the constitutional level, we usually distinguish between secondary or revision rules and first-order rules. Revision rules determine how other rules (both first- and second-order) can be revised. Because the meta-level code serves as the guardian of the algorithmic constitution its capacity to directly revise the rules should be constrained.

We propose that the meta-AI agent operate in *two modes of reflexivity*: self- and user-initiated. At the self-initiated level, the meta-agent monitors the behavior of the object-level agent, identifies violations of the rules, and intervenes with corrective action when needed. The meta-agent does not monitor the entire behavior-state of the compliance system, but focuses only on the aspects that are critical for keeping the constitutional properties of the system intact, e.g., fundamental procedural rules, such as the right of users to be notified of the imposition of sanctions and to appeal a decision that harmed their interests.[113] Although the object-level agent is not able to change the community standards explicitly, it can do so implicitly, for example, by refraining from notifying users that they were sanctioned or by blocking their attempts to send review requests. The cross-check system provides another illustration of how the object-level agent could attempt to circumvent the system's formal rules, through the creation of a list of privileged entities.[114] When a violation is detected, the meta-agent can provide immediate remedy, whether through a particularistic intervention (allowing a user to submit a review request) or by revising any changes to the object-level code that led to the problem (e.g., abolishing the cross-check bypass). This constitutional meta-monitoring requires the system designers to assemble a limited set of "fundamental violating trigger events" and to incorporate them into the meta-level code. We also assume that the meta-AI agent is not limited to the initial list and is able to revise it through a recursive learning process, in a way that should allow it cope with unforeseen loops that may undermine users' rights.

[112] See Jhaver et al., supra note 91, at 24; Inioluwa Deborah Raji, I. Elizabeth Kumar, Aaron Horowitz, & Andrew D. Selbst, *The Fallacy of AI Functionality*, *In* 2022 ACM CONFERENCE ON FAIRNESS, ACCOUNTABILITY, AND TRANSPARENCY, 959 (2022); The critique in Díaz et al., supra note 62, at 20-23.
[113] We draw here on Costantini Stefania, supra note 8, at 445.

In articulating the idea of user-reflexivity, we face the challenge of how to embed it into the meta-level algorithm. The first challenge concerns the design of the review process. As we elaborated above, in a fully AI-controlled environment, the preliminary enforcement phase and the secondary review process are conflated. We criticized this situation, noting the risks of algorithmic bias and unchecked transformative AI power. Responding to these problems requires the development of a mechanism of review that is independent of the AI algorithm that controls the base-level compliance process. The second challenge concerns the need to create reflexive mechanisms that would enable a revision of the community standards in response to systemic problems that may emerge through the review and appeal process or external pressures.

The question is how to embed these reflexive capacities in the meta-level algorithm without endowing it with too much power. As we argued above, our approach is guided by two fundamental principles: (a) uncertainty about human preferences, and consequently, about the system's overarching mission and the interpretation of the rules, and (b) the ultimate source of information about human preferences is not human behavior but human choice, as revealed through deliberation. These two principles should be algorithmically incorporated in a deliberative framework that enables the meta-level AI agent to realize the two modes of reflexivity.

There are two basic options for incorporating deliberation into the meta-level code. The first is to encode a dependence on the input of an external body, requiring the AI agent to turn to it in a predetermined list of cases. Such entities may include, for example, the Facebook Oversight Board, an out-of-court dispute settlement body certified under Article 21 of the DSA, or a multi-stakeholder entity such as the Internet Governance Forum (IGF).[115] Embedding in the code a linkage to such independent judicial-like entities provides an important counter-force to the power of the community to change the system's basic standards through externally-initiated deliberative processes.[116] The algorithmic encoding of such external dependence raises various challenges, which have been discussed in the literature on human-centered artificial intelligence. They include, for example, (a) the design of a robust and verifiable communication channel between the AI agent and the external entity; (b) allowing informative feedback between the AI agent and the external entity (which would, for example, enable the AI agent to ask for clarifications when the decision of the external entity is ambiguous); (c) constructing default rules that would determine what the AI agent should do if the input it receives does not provide an answer to the question it posed or in cases in which the external entity fails to provide an input.[117]

[114] See Policy advisory opinion on Meta's cross-check program, supra note 52.

The second option through which deliberation can be incorporated into the meta-level code is based on allowing it to directly engage with its users, relying on various deliberative procedures. Such mechanisms should allow the AI agent to resolve dilemmas by directly approaching the users, without recourse to intermediary institutions. Such procedures can be initiated either by the AI agent or by the public. In the former case, the AI agent can rely, for example, on deliberative polling, in which a representative sample of the relevant population is recruited using stratified random sampling and is then convened to deliberate and decide on a chosen problem.[118] In the latter case, the system can incorporate a petition system in which once a certain number of people join a petition, the AI agent must initiate a debate on the topic.[119]

---

[115] See, e.g., the Internet Governance Forum (IGF) and Fishkin, James S. Fishkin, Max Senges, Eileen Donahoe, Larry Diamond & Alice Siu, *Deliberative Polling For Multistakeholder Internet Governance: Considered Judgments on Access for the Next Billion*, 21:11 INFORMATION, COMMUNICATION & SOCIETY 154 (2018).
[116] See, Timothy Endicott, supra note 26, at 33 (noting the important role of judicial institutions in reconciling incommensurable considerations).
[117] See, e.g., Ben Shneiderman, *Human-centered Artificial Intelligence: Reliable, Safe & Trustworthy,* 36 INTERNATIONAL JOURNAL OF HUMAN–COMPUTER INTERACTION 495 (2020).
[118] Fishkin James, Nikhil Garg, Lodewijk Gelauff, Ashish Goel, Kamesh Munagala, Sukolsak Sakshuwong, Alice Siu & Sravya Yandamuri, *Deliberative Democracy with the Online Deliberation Platform*, The 7th AAAI Conference on Human Computation and Crowdsourcing (HCOMP 2019).
[119] This idea was inspired by the petition system of the British Parliament; see How petitions work, PETITIONS UK GOVERNMENT AND PARLIAMENT (last visited Nov. 15, 2022), https://petition.parliament.uk/help.

Crowdsourcing techniques are already used by other social media platforms such as Reddit.[120]

A key difficulty in the foregoing framework concerns the power of the AI agent to oversee the deliberation process. According to our model, the AI agent is algorithmically bound by the results of the deliberation process, but it could influence the outcome of that process by intervening in its design and dynamic. The extent of such intervention depends on the autonomy of the meta-agent. Recall that we conceptualized the meta-level algorithm as a *political satisficer,* whose goal is to satisfice human preferences, subject to deliberative constraints. The idea of political satisficing can be interpreted as allowing the AI agent to intervene in the deliberation process, if this intervention generates better understanding of human preferences. Such intervention can range from choosing the sampling technique that is used in deliberative polling processes to determining the scale and scope of the information provided to the people participating in the deliberation or to the panelists of an external out-of-court dispute settlement body (taking into account humans' limited capacity to read significant amounts of data).[121] *Thus, trying to bind the AI agent to social deliberation paradoxically gives it partial control over the very process by which it is supposed to be bound*. It is at this junction that societal and algorithmic constitutionalism collide. The challenge we face can therefore be articulated as follows: Can we design a legitimate, fair, and safe system of constitutional law that is co-governed by humans and AI algorithms?

## E. Conclusion

The idea of algorithmic constitutionalism, which we developed in this article, offers a response to the risks of algorithmic governance. It is based on three pillars: (a) layered architecture consisting of object (operative) and meta levels, (b) algorithmic meta-reasoning, and (c) correction by deliberation. Our approach is based on the thesis that the AI governance system should be infused with uncertainty about human preferences, which should consequently be translated into uncertainty about the correct interpretation of system rules and overarching mission. We argue that the system should have a variety of hardwired procedures through which it may receive input from its users about the application of the standards, the fundamental principles of the system, and potential revisions of these rules. But as we elaborated above, paradoxically, the attempt to subject the AI algorithm to external deliberative control also opens the door for the AI agent to intervene in that process, potentially undermining its very purpose. This outcome reflects a tension between societal and algorithmic constitutionalism, which we believe poses a critical new challenge to constitutional law.

[120] Marrazzo V., *The Federalists of the Internet? What Online Platforms can Learn From Reddit's Decentralized Content Moderation Scheme*, NEBRASKA LAW REVIEW BULLETIN (forthcoming 2022); Jhaver et al., supra note 91, at 15; J. Chen, B. Raghavan, P. Schmitt & T. Liu, *A Pluralist Approach to Democratizing Online Discourse*, arXiv e-prints (2021), https://pschmitt.net/docs/timbre.pdf; Facebook already partially relies on users' input, for example, in reporting abuse. See Reporting Abuse, FACEBOOK: HELP CENTER (last visited Nov. 19, 2022), https://www.facebook.com/help/1753719584844061.

[121] Meta Oversight Board decision on the cross-check system illustrates this risk. The Board criticized Meta for failing to provide it with sufficient details on the entities that receive privileged protection under the ERSR system. See Policy advisory opinion on Meta's cross-check program, supra note 52, at 27.

Our article constitutes only a first step in articulating the idea of algorithmic constitutionalism, which raises further legal, philosophical, and technological questions. We hope that it will pave the way for further investigations of this concept. A key challenge is to examine how the theoretical concept of algorithmic constitutionalism can be implemented in practice. One possible direction for research in this context concerns the idea of modular implementation. Is it possible to limit the application of algorithmic constitutionalism to those aspects of the platform's governance, which are more critical from a human-rights perspective (e.g., content moderation), and give the controlling corporation (e.g., Meta) more leeway to intervene in the regulation of less sensitive aspects (e.g., ad delivery)? Such modular implementation could provide a way to balance between the corporation's legitimate business interests and its human rights commitments, and consequently could make profit-seeking corporations such as Meta more receptive to the idea of algorithmic constitutionalism.[122]

Another potential direction for research is to examine the relations between algorithmic constitutionalism and emerging regulatory initiatives focusing on the risks of AI. Particularly relevant to this inquiry is the new European Digital Services

[122] A sharper distinction between the corporation business and human rights commitments could also be a source of extra costs. For example, the Oversight Board recommended in its cross-check opinion that Meta should prioritize expression that is important for human rights and make sure that the review of such content does not compete with Meta's business partners for limited resources. This recommendation requires Meta either to expand its review capacity or to discontinue its current practice of giving special privileges to users that have many followers (and are therefore a significant source of revenue). *Id*, at 5, 27, 37.

Act, which was approved by the European Parliament in October 2022.[123] The DSA aims to extend the responsibilities and obligations of providers of information services, which also includes digital platforms such as Facebook (which are defined as "hosting services" in the DSA).[124] One of the key contributions of the DSA is to establish a new regulatory framework for content moderation. This framework creates a semi-constitutional structure that defines the meta-procedural framework for governing digital platform content moderation actions. The structure includes, for example, provisions regarding the notification and provision of reasons (regarding, for example, removal of content, disabling access to content, or demoting content),[125] the establishment of an internal complaint-handling system that enables users to lodge complaints electronically and free of charge against a decision taken by the provider,[126] and the creation of an out-of-court dispute settlement system that would provide users with accessible redress.[127] The drafters of DSA have made a significant effort to distinguish between automated and human decision making. For example, the DSA requires providers of online platforms to ensure that the decisions made by the internal complaint-handling system "are taken under the supervision of appropriately qualified staff, and not solely on the basis of automated means," and that recipients of the service can communicate directly and rapidly with the provider in a user-friendly manner, which does not rely solely on automated tools.[128] Providers are also required to include information on the use of automated means in the notifications they send to users.[129]

---

[123] Also relevant in this context are the following legal instruments: the AI Act: European Commission, Proposal for a Regulation of the European Parliament and of the Council laying down harmonized rules on artificial intelligence (Artificial Intelligence Act) and amending certain Union legislative acts (COM(2021) 206 final); the Digital Markets Act: European Commission, Regulation of the European Parliament and of the Council on Contestable and Fair Markets in the Digital Sector (2022/1925) and amending Directives (EU) 2019/1937 and (EU) 2020/1828 (Digital Markets Act); the Machinery Regulation: European Commission, Proposal for a Regulation of the European Parliament and of the Council on Machinery Products (COM(2021) 202 final); and the Data Governance Act: European Commission, Proposal for a Regulation of the European Parliament and of the Council on European Data Governance (COM(2020). 767 final).
[124] DSA, supra note 11, at Article 3(g). For a general discussion of the DSA see, Jorge Morais Carvalho, Francisco Arga e Lima & Martim Farinha, *Introduction to the Digital Services Act, Content Moderation and Consumer Protection*, 3 REVISTA DE DIREITO E TECNOLOGIA 71 (2021), and Arno R. Lodder & Jorge Morais Carvalho, *Online Platforms: Towards An Information Tsunami with New Requirements on Moderation, Ranking, and Traceability* 537, 33(4) EUROPEAN BUSINESS LAW REVIEW (2022), Available at SSRN: https://ssrn.com/abstract=4050115.
[125] DSA, supra note 11, at Article 17(1)(a).
[126] *Id*, at Article 20(1).
[127] *Id*, at Article 21.
[128] *Id*, at Article 20(6) and Article 12(1).
[129] *Id*, at Article 16(6).

The new regulatory system created by the DSA brings to the fore the tension between algorithmic and human decision making, which has so far remained outside the reach of formal regulation, allowing digital platforms such as Facebook to be opaque about the exact role of algorithms in their content-moderation decisions. The DSA provides a unique opportunity for exploring the tension between algorithmic and societal constitutionalism in a real socio-technical environment.